\documentstyle[emulateapj,apjfonts,flushrt]{article}
\makeatletter

\@ifundefined{chapter}{\def\thebibliography#1{\section*{References\@mkboth
  {REFERENCES}{REFERENCES}}\list
  {\relax}{\setlength{\labelsep}{0em}
        \setlength{\itemindent}{-\bibhang}
        \setlength{\itemsep}{0pt}
        \setlength{\parsep}{0pt}
        \setlength{\leftmargin}{\bibhang}}
    \def\newblock{\hskip .11em plus .33em minus .07em}
    \sloppy\clubpenalty4000\widowpenalty4000
    \sfcode`\.=1000\relax}}%
{\def\thebibliography#1{\chapter*{Bibliography\@mkboth
  {BIBLIOGRAPHY}{BIBLIOGRAPHY}}\list
  {\relax}{\setlength{\labelsep}{0em}
        \setlength{\itemindent}{-\bibhang}
        \setlength{\itemsep}{0pt}
        \setlength{\parsep}{0pt}
        \setlength{\leftmargin}{\bibhang}}
    \def\newblock{\hskip .11em plus .33em minus .07em}
    \sloppy\clubpenalty4000\widowpenalty4000
    \sfcode`\.=1000\relax}}

\newlength{\bibhang}
\let\@internalcite\cite
\def\cite{\let\@citeleft(\let\@citeright)%
    \@ifstar{\citeyear}{\citefull}}
\def\acite{\let\@citeleft\relax\let\@citeright\relax%
    \@ifstar{\citeyear}{\acitefull}}
\def\citenp{\let\@citeleft\relax\let\@citeright\relax
    \@ifstar{\citeyear}{\citefull}}
\def\citefull{\def\astroncite##1##2{##1~##2}\@internalcite}
\def\citeyear{\def\astroncite##1##2{##2}\@internalcite}
\def\acitefull{\def\astroncite##1##2{##1~(##2)}\@internalcite}

\def\@citex[#1]#2{\if@filesw\immediate\write\@auxout{\string\citation{#2}}\fi
  \def\@citea{}\@cite{\@for\@citeb:=#2\do
    {\@citea\def\@citea{; }\@ifundefined
       {b@\@citeb}{{\bf ?}\@warning
       {Citation `\@citeb' on page \thepage \space undefined}}%
{\csname b@\@citeb\endcsname}}}{#1}}

\def\@cite#1#2{\@citeleft#1\if@tempswa , #2\fi\@citeright}
\def\@biblabel#1{}

\makeatother

\newcommand{\PSbox}[3]{\mbox{\rule{0in}{#3}\includegraphics{#1}\hspace{#2}}}
\newcommand{\FigNum}[1]{\unitlength 1pt \begin{picture}(55,10)(-400,35) 
                        \put(0,0){Figure #1}
                        \end{picture}}
\newcommand{\msun}{$M_\odot$} 
 
\newcommand{\persec}{\mbox{$\second^{-1}$}}
\newcommand{\percm}{\mbox{$\cm^{-2}$}}
\newcommand{\ppm}{\mbox{$\pm$}}
\newcommand{\cgsflux}{\erg\percm\persec}
\newcommand{\cgslum}{\erg\persec}

\newcommand\approxlt{\mbox{$^{<}\hspace{-0.24cm}_{\sim}$}}
\def\etal{{et~al.}}

\newcommand{\nh}{\mbox{$N_{\rm H}$}}
\newcommand{\nhtt}{\mbox{$N_{\rm H, 22}$}}
\newcommand{\ud}[2]{\mbox{$^{+ #1}_{- #2}$}}
\newcommand{\ee}[1]{\mbox{$10^{#1}$}}
\newcommand{\tee}[1]{\mbox{$\times 10^{#1}$}}

\newcommand{\perval}[2]{{#1\mbox{$^{#2}$}}} 

\def\chisqrnu{\mbox{$\chi^2_\nu$}}
\def\x1608{{4U~1608$-$522}}

\def\cenx4{{Cen~X$-$4}}
\def\aql{{Aql~X$-$1}}
\def\saxj1808{{SAX J1808.4$-$3658}}

\newcommand{\cm}{\mbox{$\rm\,cm$}}

\newcommand{\second}{\mbox{$\rm\,s$}}

\newcommand{\erg}{\mbox{$\rm\,erg$}}

\newcommand{\kpc}{\mbox{$\rm\,kpc$}}

\newcommand{\kteffinfty}{$kT_{\rm eff}^\infty$}

\newcommand{\rinfty}{\mbox{$R_{\infty}$}}

\newcommand{\chandra}{{\em Chandra\/}}

\newcommand{\rosat}{{\em ROSAT\/}}

\newcommand{\rxte}{{\em RXTE\/}}

\newcommand{\tauout}{\mbox{$\tau_{\rm outburst}$}}
\slugcomment{\it \today}
\def\ks{KS~1731$-$260}

\begin{document}

\title{Crustal Emission and the 
Quiescent Spectrum of the Neutron Star in \ks}

\author{Robert E. Rutledge\altaffilmark{1}, 
Lars Bildsten\altaffilmark{2}, Edward F. Brown\altaffilmark{3}, 
George G. Pavlov\altaffilmark{4}, 
\\ Vyacheslav  E. Zavlin\altaffilmark{5}
and Greg Ushomirsky\altaffilmark{1}
}
\altaffiltext{1}{
Theoretical Astrophysics, 
California Institute of Technology, MS 130-33, Pasadena, CA 91125;
rutledge@srl.caltech.edu, gregus@tapir.caltech.edu}
\altaffiltext{2}{
Institute for Theoretical Physics and Department of Physics, Kohn Hall, University of 
California, Santa Barbara, CA 93106; bildsten@itp.ucsb.edu}
\altaffiltext{3}{
Enrico Fermi Institute, 
University of Chicago, 
5640 South Ellis Ave, Chicago, IL  60637; 
brown@flash.uchicago.edu}
\altaffiltext{4}{
The Pennsylvania State University, 525 Davey Lab, University Park, PA
16802; pavlov@astro.psu.edu}
\altaffiltext{5}{ 
Max-Planck-Institut f\"ur Extraterrestrische Physik, D-85740 Garching,
Germany; zavlin@xray.mpe.mpg.de
}

\begin{abstract}
The type-I X-ray bursting low mass X-ray binary \ks\ was recently
detected for the first time in quiescence by Wijnands \etal, following
a \tauout$\approx$13 yr outburst which ended in Feb 2001.  We show
that the emission area radius for a H atmosphere spectrum (possibly
with a hard power-law component that dominates the emission above
3.5~keV) is consistent with that observed from other quiescent neutron
star transients, \rinfty=23\ud{30}{15 }(d/8 \kpc )~km, and examine
possible IR counterparts for \ks.  Unlike all other known transient
neutron stars (NS), the duration of this recent (and the only
observed) outburst is as long as the thermal diffusion time of the
crust.  The large amount of heat deposited by reactions in the crust
will have heated the crust to temperatures much higher than the
equilibrium core temperature.  As a result, the thermal luminosity
currently observed from the neutron star is dominated not by the core,
but by the crust.  This scenario implies that the mean outburst
recurrence timescale found by Wijnands \etal\ ($\sim$ 200 yr) is a
lower limit.  Moreover, because the thermal emission is dominated by
the hot crust, the level and the time evolution of quiescent
luminosity is determined mostly by the amount of heat deposited in the
crust during the most recent outburst (for which reasonable
constraints on the mass accretion rate exist), and is only weakly
sensitive to the core temperature.  Using estimates of the outburst
mass accretion rate, our calculations of the quiescent flux
immediately following the end of the outburst agree with the observed
quiescent flux to within a factor of a few.  In this paper, we present
simulations of the evolution of the quiescent lightcurve for different
scenarios of the crust microphysics, and demonstrate that monitoring
observations (with currently flying instruments) spanning from
1--30~yr can measure the crust cooling timescale and the total amount
of heat stored in the crust.  These quantities have not been directly
measured for any neutron star. This makes \ks\ a unique laboratory for
studying the thermal properties of the crust by monitoring the
luminosity over the next few years to decades.
\end{abstract}


\section{Introduction}

Brown, Bildsten \& Rutledge \cite*[BBR98 hereafter]{brown98} argued that
the core of a transiently accreting neutron star (NS) (for reviews of
transient neutron stars, see \citenp{chen97,campana98b}) is heated by
nuclear reactions deep in the crust during the accretion outbursts. The
core is heated to a steady state in $\sim$\ee{4} yr (see
\citenp{colpi00} for a detailed calculation), after which the NS emits a
quiescent thermal luminosity (BBR98)
\begin{equation}
\label{eq:brown}
L_q = 8.7\times 10^{33} \left(\frac{\langle \dot{M} \rangle}{10^{-10} 
M_\odot {\rm yr}^{-1}}\right) \frac{Q}{1.45 {\rm MeV}}\; {\rm 
erg \; s}^{-1}, 
\end{equation}
where $\langle \dot{M} \rangle$ is the time-averaged (over the NS core
thermal timescale) mass-accretion rate onto the NS, and $Q$ is the
amount of heat deposited in the crust per accreted nucleon
(\citenp{haensel90}; see \citenp{bildstenrutledge00} for a review).  For
an accretion flux onto the NS ($F_{\rm accretion}=\epsilon \dot{M}
c^2/(4\pi D^2)$), the ``rock bottom'' quiescent flux due to deep crustal
heating is then
\begin{equation}
\label{eq:fq}
F_q \approx \frac{\langle F_{\rm accretion} \rangle}{135} \,
\frac{Q}{1.45 {\rm MeV}} \, \frac{0.2}{\epsilon},
\end{equation}
where $\langle F_{\rm
raccretion}\rangle\equiv\epsilon\langle\dot{M}\rangle c^2/(4\pi D^2)$
is the accretion flux averaged over the NS core thermal timescale, and
the accretion efficiency $\epsilon=0.2$ for accretion luminosity of
$(GM/R)\langle \dot{M}\rangle$.

Since $\langle\dot{M}\rangle=\dot{M}_{\rm outburst} (\tau_{\rm
outburst}/\tau_{\rm rec})$, where $\tau_{\rm outburst}$ is the mean
outburst duration and $\tau_{\rm rec}$ is the mean recurrence
timescale, this scenario relates the quiescent luminosity to the
outburst properties; comparisons to observations of several quiescent
neutron stars (qNSs) match the predictions reasonably well
\cite{brown98,rutledge00}.  It also helps us to understand why the
black hole systems in quiescence are so much fainter than NS systems
\cite{garcia01}, as they cannot be thermally emitting.
Equation~(\ref{eq:fq}) is the minimum quiescent flux from the neutron
star.  Should residual accretion occur in quiescence
\cite{menou99,narayan01}, the accretion luminosity would be in
addition to the thermal emission already present.

\ks\ is a transient, type-I X-ray bursting neutron star approximately
3.8 degrees from the Galactic center \cite{jvp95}.  It was discovered in
outburst in 1989, when its luminosity was 1.3\tee{37} (d/8 \kpc)$^2$
\cgslum ; subsequent analysis found that it had been in outburst at
least since Oct 1988 (\citenp{sunyaev89,sunyaev90}), and has remained
bright since then (see \citenp{wijnands01b} for a complete description
of observations). Sunyaev et al. (1990) suggested that it was a
transient, apparently because it had not previously been detected,
although no strong upper limits on previous X-ray emission were
given. (This was noted later by \acite{barret98}, who suggested the
object may not be a transient at all. We adopt an outburst timescale of
$\tau_{\rm outburst}\approx$ 13 yr, although the outburst may have been
longer). Type I X-ray bursts were observed, at the rate of $\sim$10 per
day, and studied in detail with RXTE \cite{muno00}.  Nearly coherent
oscillations (580 Hz) have been observed during the type-I bursts, and
the frequency is interpreted as the spin frequency of the NS
\cite{smith97}.  The uncertainty in the localization of this source was
initially 4.2\arcmin\ (90\% confidence).  An improved error circle
(10\arcsec\ with \rosat/HRI; \citenp{barret98}) found 13 candidate IR
counterparts down to $J=15.4$ (10~$\sigma$).  The integrated HI 
column density in the direction of \ks\ is \nhtt=0.35
\footnote{W3nH, at http://heasarc.gsfc.nasa.gov; \citenp{dickey90}}
($N_H=$\nhtt\ee{22} \perval{cm}{-2}).  \acite{barret98} found
\nhtt=1--6 ($A_V$=5--34; \citenp{predehl95}), suggesting that \nh\ is
variable, and thus there must be significant contribution to \nh\ from
the system itself.

An \rxte/PCA scan of the Galactic Center found that \ks\ had entered a
low luminosity state \cite{wijnands01b} in early 2001.
\acite{wijnands01b} then observed the source with \chandra/ACIS-S in
late March ($\sim$1 months after the end of the outburst) and detected
it at an unabsorbed flux of $\sim
2\tee{-13}{\rm\,erg\,cm^{-2}\,s^{-1}}$ (0.5--10~keV).

\subsection{Crust Thermal Emission}
\label{sec:crust-therm-emiss}

The scenario described by equation (\ref{eq:fq}) applies when the
neutron star crust and atmosphere resemble that of a cooling neutron
star -- that is, the temperature decreases with increasing radius.  For NS
transients with short outbursts, such as Aql~X-1 ($\tau_{\rm
outburst}\sim 30{\,\rm d}$), this is a good approximation, as the
increase in crust's temperature from the heating during the outburst is
small \cite{ushomirsky01}.  This is not the case for \ks.  The
duration of its most recent outburst is of the order of (or longer
than) the thermal diffusion timescale in the crust (BBR98).  In a
sense, \ks\ in quiescence resembles a neutron star that accreted
steadily at the \emph{outburst} accretion rate, except that the core
temperature will be at the lower value set by the time-averaged
accretion rate (over the previous $10^{4}{\,\rm yr}$).

To illustrate how the crust is so dramatically heated during a long
outburst, consider the rise in temperature if no heat were conducted
away from the crust during the outburst.  As we describe in
\S~\ref{sec:crust}, the total heat capacity of the region of the crust
where most of the heat is deposited is $C\sim
5\times10^{35}$~erg~K$^{-1}$.  During the outburst, the total amount of
heat deposited is $Q\dot{M}_{\rm outburst}/m_p\approx
10^{44}\,{\rm erg}$ and so, {\it if no heat is conducted away from
the heating region}, the temperature there can rise to $2\times10^8$~K
during the 13~year outburst.  Even when thermal conduction is taken into
account, the rise in crust temperature can still be $>
5\tee{7}\,{\rm K}$, which is the typical temperature the crust and
core would have if \ks\ accreted steadily at
$\langle\dot{M}\rangle=\dot{M}_{\rm outburst} (\tau_{\rm
outburst}/\tau_{\rm rec})\approx 2.6\times10^{-11} M_\odot$~yr$^{-1}$
(for $\tau_{\rm rec}\approx1500$ yr; see \S~\ref{sec:core}). 

 If the crust, which is composed of the ashes of H/He
burning, has a low thermal conductivity, then there is a substantial
temperature gradient between crust and core.  As discussed by
\acite{brown00} for the case of steadily accreting neutron stars, when
there is a substantial thermal gradient in the inner crust, the
temperature of the crust reaction layers becomes decoupled from that
of the core.  

For the NS in \ks, this means that, until the crust has thermally
relaxed, we cannot directly infer its core temperature, as is possible
for short-outburst transients such as \aql.  However, this also means
that one can directly compute the current quiescent luminosity from
\ks\ using the fluence during the last outburst, as the thermal state
of the crust is only weakly sensitive to the temperature of the core,
and, hence, is nearly independent of the uncertainty in the accretion
history over the past $\sim10^4$~yrs (the thermal time of the core),
unlike in the case of short-outburst transients.  Our simulations,
presented here, predict a quiescent luminosity which agrees with the
observed value to within observational and theoretical uncertainties 
(factor of a few) for the fiducial value of $Q$ and outburst fluence.
Moreover, this dependence of the thermal flux in the crust-dominated
regime on $\dot{M}_{\rm outburst}$, and not on
$\langle\dot{M}\rangle$, allows us to predict the evolution of $L_q$,
which can be directly confronted with observations.

The domination of the quiescent luminosity by the cooling crust sets
\ks\ apart from the other neutron star transients with short duration
outbursts and makes it an ideal laboratory for separately measuring
the thermal properties of the crust.

\subsection{Outline of the Paper}\label{sec:outline}

In this paper we report our analysis of the \chandra\ observation
\cite{wijnands01b} of \ks\ in quiescence and describe our simulations
of the thermal state of its crust and core.  We begin in
\S~\ref{sec:anal} by describing the \chandra\ observation and spectral
analysis of \ks.  We include a description of possible IR counterparts
(\S~\ref{sec:ir}).  After describing the observations, we then discuss
the implications for the crust and core thermal structure as outlined
above.  We first apply (\S~\ref{sec:core}) the scenario in which the
crust heating for any single outburst is negligible, i.e., we treat
\ks\ as if it were a short-duration outburst, such as Aql~X-1.  This
analysis has been done previously by \acite{wijnands01b}.  We then
explain, in \S~\ref{sec:crust}, why this analysis is inapplicable to
this source, and how the quiescent luminosity is determined by the
thermal properties of the crust (its thermal diffusion timescale and
heat capacity) for the next 1--30~yr.  We present simulations of the
quiescent lightcurve and demonstrate that monitoring observations can
constrain these properties, similar to proposals for glitching pulsars
\cite{riper91,chong94,hirano97,cheng98}.  We conclude in
\S~\ref{sec:con} with a summary and discussion of these results.

\section{\chandra\ Observations}
\label{sec:anal}

The data were obtained from the \chandra\ public archive. The
discovery and observations have been previously analyzed by
\acite{wijnands01b}, and details of the trigger and history of \ks\
are included there. The observation was triggered when an \rxte/PCA
scan of the Galactic Center found that \ks\ had entered a low
luminosity state \cite{wijnands01b}, and was made 2001 Mar 27
00:18-06:23 (TT) with the ACIS-S instrument \cite{chandra} for a total
exposure time of 19401 seconds. The X-ray source was imaged on the S3
(backside illuminated) chip, which was operated as a 1/4 array with
0.8 sec exposures. We analyzed the data using the CIAO v2.1
\footnote{http://asc.harvard.edu/ciao2.1/} with CALDB v2.6 and XSPEC
v11 \cite{xspec}. 

Two X-ray point sources separated by 31.3\ppm0.1\arcsec\ are found
with {\em celldetect} above a signal/noise ratio of 5, listed in
Table~\ref{tab:objs}.  We compared this image with an archived
\rosat/HRI image of this region
(rh400718)\footnote{http://heasarc.gsfc.nasa.gov} taken in 1997 when
\ks\ was in outburst. The position of \ks\ had been previously
determined this way \cite{barret98} and clearly \ks\ is at the
position of \chandra\ source \#1.  

X-ray source counts were extracted within an area 5 pixels in radius
about the \ks\ source position, for a total of 183 counts. At 0.0075
counts/frame, the pileup fraction is negligible.  Background was taken
from an annulus centered on the source position, with radii of 10 and
80 pixels. The expected number of background counts in the source
region is 3. A KS test \cite{press} finds that the times of arrival
(TOA) of the 183 counts is consistent with a constant countrate.
We place 3$\sigma$ upper-limits on variability of
$3\times1./\sqrt{183 {\rm \; counts}}$=22\% rms. We
also produced a power-density spectrum to search for a pulsed signal
(number of frequency bins: 12750, with frequency resolution of
5\tee{-5} Hz, and a Nyquist frequency of 0.625 Hz; see
\citenp{press}), using barycentered TOAs (tool {\em axbary}).  No
evidence for a pulsed signal is found: the largest Leahy-normalized
power \cite{leahy83} was 20.37 (with a probability of chance occurrence
from a Poisson-distributed countrate of 0.48).  The absence of a
coherent signal is not surprising in light of the detection of
pulsations during type-I X-ray bursts \cite{muno00} at $\sim$580 Hz,
well above our Nyquist frequency. 

\subsection{Spectral Analysis of \ks} 
\label{sec:spectral}

We binned the \ks\ data into 10 PI bins (0.5-10.0 keV), and fit
several single component spectral models (powerlaw, H atmosphere, or
Raymond-Smith, a multicolor disk and blackbody).  Galactic absorption
is initially left as a free parameter. The best-fit models were all
statistically acceptable; the parameters are given in
Table~\ref{tab:fits}.

While all models we investigated are statistically acceptable, some
can be argued against on physical grounds.  The pure power-law
spectrum is unusually steep ($\alpha$=5.2\ppm0.6), typical more of a
thermal spectrum than other processes. The emission measure $\int
n_e\, n_h dV$ from the Raymond-Smith spectrum is higher by 2 orders of
magnitude than typical from active stellar coronae in the analogous RS
CVn systems (\citenp{dempsey93a,dempsey93b}; see
\citenp{bildstenrutledge00} for discussion on coronal emission from
companions in X-ray transients).  The disk black-body spectral model
implies an inner disk radius of $\sim$0.7 km -- considerably smaller
than a neutron star.  Our best-fit blackbody spectrum is consistent
with that found by \acite{wijnands01b}.

As pointed out by BBR98, for accretion rates \approxlt 2\tee{-13}
\msun\perval{yr}{-1}, gravity stratifies metals in the NS atmosphere
faster than they can be provided by accretion (Bildsten, Salpeter, \&
Wasserman \citenp*{bildsten92}), making a pure H atmosphere the
appropriate description of the NS photosphere in quiescence.  Due to
the strongly energy-dependent opacity of free-free transitions, these
spectra are significantly different from black-bodies
\cite{rajagopal96,zavlin96}.  Unlike the results from black-body fits,
the inferred NS radii (\rinfty=$r/\sqrt{1-2GM/(rc^2)}$ where $r$ is the
proper radius, and $\infty$ means the value as measured by a distant
observer) of the four observed field transients using H atmosphere
spectra are consistent with what are  expected theoretically:
\rinfty$\sim$12 km (\cenx4, \citenp{rutledge01}; \aql,
\citenp{rutledge01b}; 1608-522 \citenp{rutledge99}; and 4U 2129+47
\citenp{rutledge00}), thus supporting the notion that much of the
emission originates from a neutron star surface.  

The quiescent spectrum for qNSs is thus interpreted as thermal
emission from a pure Hydrogen atmosphere NS, possibly with an
underlying power-law (whose origin is not understood) which dominates
the spectrum at high ($>$3 keV) energies (see Rutledge \etal\
\citenp*{rutledge01,rutledge01b}).  The spectral fit for a H
atmosphere spectrum alone (no power-law) gives \kteffinfty=120\ppm30
eV, an emission area radius of \rinfty=10\ud{10}{5} (d/8 kpc) km, and
an unabsorbed flux of 1.8\tee{-13} \cgsflux (0.5-10 keV; the absorbed
flux is 3.8\tee{-14} \cgsflux), corresponding to a luminosity of
$L_X=$1.4\tee{33} (d/ 8\kpc)$^2$ \cgslum (0.5-10 keV).  The 90\%
confidence range in the unabsorbed flux is (1.1--4.6)\tee{-13}
\cgsflux.  The best-fit spectrum is shown in Fig.~\ref{fig:ks1731}.
The bolometric thermal luminosity (as observed at infinity) is $L_{\rm
bol, \infty}=4\pi\rinfty^2 \sigma (T_{\rm eff}^\infty)^4=$2.7\tee{33}
(d/8 \kpc)$^2$ \cgslum (uncertain by a factor of $\sim$3, due largely
to spectral uncertainty).  If 100\% of the emergent luminosity were
from accretion (at an efficiency of 0.2), this sets $\dot
M=2.3\tee{-13}$ \msun\perval{yr}{-1}, sufficiently low for the
assumption of an H atmosphere.

When we include an underlying power-law with all parameters free, the
S/N of the spectrum does not constrain the model parameters to better
than an order of magnitude. When we hold \rinfty\ and $\alpha$ fixed
at typical values (\rinfty=12.5 km for d=8 kpc; $\alpha=0.85$;
cf.~\citenp{rutledge01,rutledge01b}), the spectral model provides an
acceptable fit to the data. This demonstrates that the spectrum of
\ks\ in quiescence is consistent with that observed from other qNSs.

\subsection{Possible Infrared Counterparts}
\label{sec:ir}

A comparison between Chandra astrometry and 2MASS images finds a
possible IR counterpart at the position of source \#2 (see below) but
not of \ks, with limits in J, H and K$_s$ of 15.8, 15.1 and 14.3
magnitudes respectively
\footnote{http://www.ipac.caltech.edu/2mass/overview/about2mass.html}.
Comparing with IR sources previously examined as possible counterparts
to \ks\ \cite{barret98}, the closest object is source ``H'', which is
1.5\ppm1.1\arcsec\ away (we adopt 1\arcsec\ positional uncertainty for
\chandra, and 0.5\arcsec\ uncertainty for the IR source).  The second
closest is ``G'', which is 3.3\arcsec\ away. In addition, there is a
fainter IR object present in the H-band image (``new'' object), but
not in the J-band image taken by Barret, which is not marked, but
which is consistent with the \chandra\ position for \ks.  Further work
in the IR -- for example, searching for ellipsoidal variations in
quiescence -- is needed for a positive identification of the
counterpart.

Source \#2 is consistent in position with 2MASS J173412.7-260548
($r=0.46$\arcsec\ in distance, about the accuracy of the 2MASS position
reconstruction error).  For an IR field source density of
$\rho$=6.9\tee{-3} arcsec$^{-2}$ (217700 objects in the 2MASS catalog
within a 1000\arcsec\ radius)
the probability of a chance coincidence is prob=$1-\exp(-4\pi r^2
\rho)$=3.9\%, which is the significance of the association of the X-ray
and IR objects.  The source's 2MASS magnitudes and colors are
J=12.264\ppm0.033, $J-H$=0.948 and $J-K_s$=1.33.

When this work was largely complete, \acite{wijnands01c} used relative
astrometry with 2MASS, and found that the \chandra\ localization is
coincident with the ``new'' object and not ``H'' and ``G''.  

\section{Core Dominated Emission, Crust Dominated Emission, and the
Future Lightcurve of \ks}

Having described the observations and spectral analysis of \ks, we now
turn our attention to the thermal properties of the neutron star's
crust and core.  We begin by estimating the amount of mass deposited
during the previous outburst; we then describe what the thermal
structure would be if the change in crust temperature during this
outburst were small (BBR98; \citenp{wijnands01b}).  This is the case
for transients with short-duration outbursts.  Because of the length
of the outburst for this source, however, the crust is substantially
heated; \S~\ref{sec:crust} describes how the observed emission is
determined by the heating of the crust during the past outburst, and
not by the core equilibrium temperature.  Using the method of UR01, we
simulate the evolution of the quiescent lightcurve for different
regimes of the crust and core microphysics.  

\subsection{The Outburst Fluence}
\label{sec:outburst-fluence}

\ks\ has been observed several times, always in outburst, since 1988,
and nearly continuously so since Jan 1996 with the
\rxte/ASM\footnote{http://xte.mit.edu}.  To estimate $\langle F_{\rm
outburst} \rangle$, we integrated the ASM counts using only data in
which there was a $>3\sigma$ 1-day detection and taking only 1 day
detections in which the previous observation was made $<$4 days (to
insure adequate sampling) prior.  Periods in which the time to the
previous observation was $>$4 days were not included.  The total
fluence is 13660 ASM c/s$\times$days, over an integration of 1599 days
(including both detections and non-detections).  If we assume an
additional countrate at the 2$\sigma$ level on those observations when
the countrate was below 3$\sigma$ detection, the total fluence would
only increase by 7\%, which is our 2$\sigma$ upper-limit for the
uncertainty in integrated counts.  Assuming a $kT_{\rm bremss}$=5 keV
bremsstrahlung spectrum with \nhtt=1.0 (which gives 6.5\tee{-10}
\cgsflux per 1 ASM c/s, corrected for absorption, 0.01--20 keV, from
W3PIMMS \footnote{http://heasarc.gsfc.nasa.gov}; in the 2--10 keV
range, the conversion factor is 3.0\tee{-10} \cgsflux per 1 ASM c/s ),
we find a mean outburst flux, corrected for absorption, of $13660 \,
{\rm ASM\; c/s}\times {\; \rm d}/1599 {\rm \; d} \times$ (6.5\tee{-10}
\cgsflux\ per 1 ASM count) = 5.6\tee{-9} \cgsflux .  We conservatively
estimate the spectral uncertainty to be at the 40\% level (for a
change in $kT_{\rm bremss}$ between 2 keV and 8 keV, the ASM
counts/flux conversion changes by 40\%).  Extrapolating this mean to
the entire outburst, we find an outburst fluence of ${\cal F} =
5.6\tee{-9}\cgsflux \times (13{\,\rm yr}\times 3.2\tee{7}{\rm\,s/yr})
= 2.3{\,\rm erg\,cm^{-2}}$.  This estimate is greater by a factor of
$\sim$6 than that of \acite{wijnands01b}, who estimated ${\cal F}$
using the minimum observed outburst flux and an outburst duration of
11.5~yr, rather than 13~yr.

\subsection{A First Estimate: Core-Dominated Emission}
\label{sec:core}

The quiescent bolometric thermal flux is
$F_q=3.5\times 10^{-13}\cgsflux$.  If the change in crust temperature
during the outburst were small, then equation~(\ref{eq:fq}) would apply:
the time-averaged accretion flux is $\langle F_{\rm accretion}\rangle =
{\cal F}/\tau_{\rm rec}$, so that
\begin{equation}
\label{eq:rec}
   \tau_{\rm rec}\approx \frac{{\cal F}}{135 F_q}
   \frac{Q}{1.45{\,\rm~MeV}} \frac{0.2}{\epsilon} = 1500{\rm\,yr}.
\end{equation}
There is an uncertainty of about \ppm0.5 dex in this value, mostly due
to the uncertainty in the bolometric value of $F_q$.  This estimate is
similar to the one arrived at independently by \acite{wijnands01b},
although our estimated mean recurrence time is longer (1500 \ppm 0.5
dex yr vs.\ 200 yr), since our time-averaged outburst flux is a factor
of six greater than that assumed by \acite{wijnands01b}.

Strictly speaking, this estimate of $\tau_{\rm rec}$ is the recurrence
time to an outburst of comparable fluence.  A month-long outburst would
have just 1\%\ of the fluence of the previous outburst and would not
affect the estimate of eq.~(\ref{eq:rec}).  This estimate also
neglects any neutrino emission, which reduces the effective value of
$Q$ and hence $\tau_{\rm rec}$.  More seriously, however, this
analysis ignores the fact that the crust is strongly perturbed away
from the core equilibrium temperature during such a long outburst.  As a
result, the quiescent emission currently observed is set by the
thermal relaxation of the heated crust, and as we now demonstrate, is
mostly decoupled from the thermal state of the core.

\subsection{Crust Dominated Emission}
\label{sec:crust}

During an outburst, the heating from the reactions raises the
temperature in the crust around the reaction layers.  After the
outburst ends, the crust thermally relaxes and the thermal profile
comes to resemble a cooling neutron star, i.e., the temperature
decreases with radius.  For neutron star transients with ``typical''
outburst fluences and recurrence times (that is, $\tau_{\rm
outburst}\sim 1$~month, with an outburst luminosity $\sim 0.1$~
Eddington), the variations in the crust temperature and hence $L_{q}$
are small (BBR98; UR01).  For example, UR01 found that $L_{q}$ varied
by $\sim$~few\% for $\tau_{\rm rec}=1$~yr and by $\sim$~30\% for
$\tau_{\rm rec}=30$~yr.  The variation is small because the amount of
energy deposited in such short outbursts is not significant compared
to the heat content of the crust, and so the crust is not heated
substantially compared to its temperature when in thermal equilibrium
with the core.  

As our order of magnitude estimate of the crustal temperature for \ks\
in \S~\ref{sec:crust-therm-emiss} demonstrates, the situation is completely
different in the case of \ks, with its $\tau_{\rm outburst}\approx
13$~yr outburst.

To ascertain the magnitude of the deviation of $L_q$ from the value
predicted by Eq.~(\ref{eq:brown}), we performed several simulations of
thermal relaxation of the NS crust in a transient with observational
properties ($\tau_{\rm outburst}$, $\tau_{\rm rec}$, $F_{\rm
outburst}$) inferred for \ks\ using the methods and microphysics as
described in UR01.  In summary, we solve the non-relativistic heat
equation in the crust, from $\rho\approx10^8$~g~cm$^{-3}$ to
$1.5\times10^{14}$~g~cm$^{-3}$.  Heating due to nuclear reactions in
the deep crust is simulated by depositing energy at densities
corresponding to the nuclear transitions computed by
\acite{haensel90}, with the amount of energy set by the instantaneous
accretion rate.  At the outer boundary, we use the flux-temperature
relation for a fully accreted crust from
Potekhin~\etal~\cite*{potekhin97}.  The core is taken to be isothermal
(a good assumption for the timescales of interest because of the
large thermal conductivity), and its temperature evolves according to
the mismatch between its neutrino emissivity and the flux from the
crust.  We start our simulations with the temperature profile
corresponding to persistent accretion at the rate
$\langle\dot{M}\rangle$ corresponding to the time average of
$\dot{M}_{\rm outburst}$ over the recurrence time.  We then evolve the
model through several outburst/quiescence cycles, as necessary for the
model to ``forget'' the initial conditions and reach a limit cycle.

If the crust resembles that of a neutron star steadily accreting at
$\dot{M}_{\rm outburst}$, then the estimate of the recurrence time
(eq.~[\ref{eq:rec}]) is inapplicable (except as a lower limit).  This
is because the crust temperature is decoupled from that of the core.
A full survey of parameter space should include simulations over a
range of $\tau_{\rm rec}$.  This is beyond the scope of this initial
paper; rather, we presume that $\tau_{\rm rec}$ is given by
eq.~(\ref{eq:rec}) in order to survey the influence of the crust and
core microphysics.

In these simulations, we examine the two main uncertainties in NS
microphysics, (1) the impurity fraction, and, hence, the conductivity
of the crust and (2) the possible presence of ``enhanced'' core
neutrino cooling mechanisms,\footnote{Neutrino emission due to crustal
bremsstrahlung is insignificant at the relevant temperatures, but is
included in all models.} such as direct Urca or pion condensation.
First, the composition of accreting NS crusts is set by the nuclear
processing of products of burning in the upper atmosphere.  Published
calculations \cite{sato79,haensel90} presume that burning proceeds to
pure iron, and, hence, the crust has no impurities.  However, recent
work by Schatz and collaborators \cite{schatz99,schatz01} shows that
burning products are a mix of elements substantially heavier than
iron, with the impurity parameter\footnote{The impurity parameter is
the deviation of the nuclear charges $Z$ of the elements of the mix
from the average charge $\langle Z\rangle$, $Q_{\rm imp}=\langle
(Z-\langle Z\rangle)^{2}\rangle$, where the average takes into account
the relative abundance of the species.} $Q_{\rm imp}$ comparable to
the square of the average nuclear charge $\langle Z\rangle^2$.  We
therefore set bounds on the crustal conductivity by using electron-ion
scattering \cite{yakovlev80:_therm}, which has the same form as
electron-impurity scattering with $Q_{\rm imp}=Z^2$, and
electron-phonon scattering \cite{baiko95}, which is appropriate for a
pure crystal.  We refer to these two cases as ``low $K$'' and ``high
$K$'', respectively.  Second, the uncertainty in core cooling is
covered by simulations with ``standard cooling'' and ``enhanced
cooling'' (modified Urca appropriately suppressed by nucleon
superfluidity, and neutrino emission as when the core is a pion
condensate, respectively; see UR01 for details).

To illustrate the deviation of the crustal temperature from the
equilibrium core temperature and the subsequent thermal relaxation of
the crust, we show in Fig.~\ref{fig:tevol} the time evolution of the
crustal temperature in the low $K$, standard cooling case for
observational parameters inferred for \ks\ ($\tau_{\rm rec}=1500$ yr,
$\tau_{\rm outburst}=13$ yr, and $\dot{M}_{\rm
outburst}=3\times10^{-9} M_\odot$~yr$^{-1}$).  Vertical slices of the
surface in the density-temperature plane show the instantaneous
temperature distribution in the crust, while slices in the
time-temperature plane show the time evolution of the temperature
since the beginning of the outburst.  Clearly, the crust is heated to
well above its pre-outburst temperature profile, which tracks the core
temperature (note that, because the crust is heated to well above its
temperature in equilibrium with the core, this qualitative result is
not very sensitive to the assumed $\tau_{\rm rec}$).  A steep
temperature gradient carries most of the heat from the deep crustal
heating region around $10^{12}\mbox{--}10^{13}$~g~cm$^{-3}$ into the
core, while a shallower temperature gradient carries a fraction of the
heat towards the surface.  The temperature near the top of the crust
($\sim$\ee{10} g \perval{cm}{-3}; and, hence, the X-ray luminosity,
$L_q\propto T_{\rm top}^{2.4}$; \citenp{potekhin97}) reaches a maximum
at the end of the outburst (time=13~years), and then decays back to
the pre-outburst value on the thermal timescale of the crust
($\sim30$~yrs for this model).  Thermal evolution in other cases is
similar (but, of course, is different in the absolute magnitude of the
temperature change), and the temperature changes are always much
bigger than those that occur during short-duration outbursts (cf.\
Fig.~1 of UR01).

In Fig.~\ref{fig:greg}, we show the long-term evolution of the
quiescent surface luminosity from \ks\ after accretion onto the
compact object has ceased for the four cases discussed above, namely,
low and high crustal conductivity $K$, and standard and enhanced core
neutrino emission (see the legend on the plot).  First consider
Fig.~\ref{fig:greg}a, where we set $Q=1.45\, {\rm MeV}$.  In all four
cases, the transition from crust-dominated (at early times) to
core-dominated cooling (at late times) is evident as a drop in the
luminosity, ranging from 50\% (high $K$, standard cooling case) to a
factor of $100$ (low $K$, enhanced cooling case).  Regardless of the
core neutrino emissivity, this transition occurs at $\sim30$~yrs for
low conductivity crusts, and after $\sim1$~yr for high conductivity
crusts.  These timescales are easy to understand, as they are just the
thermal time to the appropriate depth in the crust (cf.  Figure 3 of
UR01).  Since the low $K$ crust has a longer thermal time, it stays
hot about 30 times longer than the high $K$ crust.

After the crust thermally relaxes, $L_q$ is set by the emission from a
hot core.  If the only core neutrino emission mechanism is modified
Urca, suppressed by nucleon superfluidity, then the amount of heat
lost by neutrinos from the core is negligible.  Therefore, all the
heat deposited into the star comes out as thermal emission, and $L_q$
asymptotically approaches the value given Eq.~(\ref{eq:brown}), with
$Q\approx 1.45 {\rm MeV}$ (see solid and dashed lines in
Fig.~\ref{fig:greg}).  On the other hand, when enhanced neutrino
cooling is allowed, $\sim 90$\% of the deposited heat escapes as
neutrinos (\citenp{colpi00}; UR01), and only the remaining $\sim10$\%
is radiated thermally from the surface.  In this case, the
``effective'' value of $Q$ in Eq.~(\ref{eq:brown}), in the sense of
the energy retained in the star and re-radiated thermally, is
$Q\approx 0.1{\rm MeV}$.

As shown above, the crust is heated to well above the core temperature
during the long outburst.  How does the temperature difference between
the crust and the core depend on the crustal conductivity? As an
estimate, we can assume that, during the outburst, the crust comes to
an equilibrium where the heat deposited around
$\rho\approx10^{12}-10^{13}$~g~cm$^{-3}$ at a rate $\dot{E}$ is
conducted away from this region. This is a very good assumption for
the high conductivity case, and an acceptable one for the low
conductivity case, where the crustal thermal time is comparable to the
outburst duration.  During the outburst, most of the heat is conducted
into the core with a flux $F\approx K(T_{\rm crust}-T_{\rm
core})/\Delta r$, where $\Delta r \approx 500$~m is the distance
between the crustal heating region and the core.  Thermal balance then
requires $\dot{E}/4\pi R^2\approx K (T_{\rm crust}-T_{\rm
core})/\Delta r$, or $T_{\rm crust}-T_{\rm
core}\approx1.2\times10^8$~K~$/K_{19}$, where $K_{19}$ is the
conductivity in units of
$10^{19}$~erg~cm$^{-1}$~s$^{-1}$\perval{K}{-1}.  In the low
conductivity case, $K_{19}\sim1$, so $T_{\rm crust}\gg T_{\rm
core}$. We see from Fig.~\ref{fig:greg}a that $L_q$ at early times in
the low $K$ case is the same for standard and enhanced cooling (solid
and dotted lines). This is now easy to understand, since, as we showed
above, the crustal temperature is decoupled from the core temperature
if the conductivity is low.  This situation is exactly the same as for
persistent accreters discussed by Brown~\cite*{brown00}.  On the other
hand, for high conductivity crusts, $K_{19}\gtrsim 10$, so $T_{\rm
crust}$ can not significantly deviate from $T_{\rm core}$ during the
outburst. Therefore, crust-dominated $L_q$ in the high $K$ case
(dashed and dash-dotted lines in Fig.~\ref{fig:greg}a) more closely
reflects the core temperature, which is quite different between the
standard and enhanced cooling cases.

From examination of lightcurves in Figs.~\ref{fig:greg}a, it is likely
that \ks\ was observed in quiescence during the crust-dominated phase,
and is presently evolving through this phase.  In three of four cases
considered in Fig.~\ref{fig:greg}a, the predicted quiescent luminosity
$L_q$ right after the outburst is a factor of 3 higher than the observed
$L_{\rm bol}$ (itself uncertain by a factor of 3, due to spectral
uncertainty). Since $L_q$ depends directly on $Q\int \dot{M}_{\rm
outburst} dt$ over the outburst duration, this difference could also be
due to a systematic overestimate of $\dot{M}_{\rm outburst}$ during the
$\approx13$~yr outburst, only the last 5~years of which have been
covered $\sim$~daily with RXTE/ASM.  Alternatively, this difference
could be due to a value of $Q$ different from the fiducial, since the
exact value of $Q$ is uncertain (see Schatz \etal\ 1999, 2001
\nocite{schatz99,schatz01} for a discussion on different crustal
compositions than assumed by \citenp{haensel90}).  To provide
lightcurves which extrapolate from the currently observed luminosity, in
Fig.~\ref{fig:greg}b we adjusted $Q$ (or, equivalently, $\int
\dot{M}_{\rm outburst} dt$) by factors of 0.3 to 2, as noted in the
figure caption, to obtain $L_q\sim2.7\times10^{33}$~erg~s$^{-1}$ at the
start of quiescence.

We stress that the overall normalization of these lightcurves is
uncertain due to uncertainty in $Q$ and outburst fluence, but the
shape is determined by the microphysics of the crust and the
temperature of the core.  The time evolution of the luminosity will
permit distinction between the four cases, and, hence, directly infer
the integrated conductivity of the crust and possibly indicate whether
enhanced neutrino emission mechanisms are operating in the core.  In
particular, observing the drop in the luminosity will permit the
measurement of the thermal timescale of the crust.  In addition, the
relative magnitude of the drop will tell us about the presence or
absence of enhanced neutrino emission from the core.  Such
measurements (accurate to \approxlt 10\%) are well within the
capabilities of present X-ray instrumentation, as the present
detection demonstrates. Observations over the next year will be able
to exclude the case where the crust is a pure crystal and hence has
very high conductivity.

Since the initial understanding of Type I bursts as thermonuclear
flashes, there has been an open question as to how much flux is rising
up from deep parts of the star into the burning region (see
\citenp{bohdan83,bildsten95}).  Large values of this flux ($\sim 1$
MeV per accreted nucleon) can stabilize burning at accretion rates
lower than otherwise expected.  \acite{brown00} first showed that most
of the crustal heating in a persistent source went into the NS core
and that, at most, 100 keV per accreted nucleon came out into the
upper parts of the star.  The quiescent observations of \ks\ shortly
after outburst strongly support this picture and tell us that deep
nuclear heating will not impact the Type I X-ray burst behavior.

Assuming that the source has been transiently accreting for the past
$10^{4}{\rm\,yr}$ or so, with outburst fluences similar to that of the
most recent observed outburst, we can constrain the core temperature to be
less than the maximum crust temperature.  To determine the most
conservative (i.e., largest) upper bound on this peak crust temperature,
we use the following assumptions.  We take the atmosphere to be composed
of pure iron (which has a higher opacity than hydrogen/helium) and we use
low thermal conductivity (electron-ion scattering) in the crust.  These
assumptions lead to the largest temperature rise between the photosphere
and the crust.  Taking $kT_{\rm eff}=130\rm\, eV$, we find a peak crust
temperature, and hence a maximum core temperature, of $3.5\tee{8}\rm\,K$.

\section{Discussion and Conclusions}
\label{sec:con}

The large amount of heat deposited during the recent (and the only
observed) outburst of \ks, and the lengthy duration of the outburst
compared to the thermal diffusion time of the NS crust, imply that the
quiescent flux recently observed is dominated by emission from the
cooling crust and not the core.  The importance of this result is that
the quiescent luminosity {\em is calculated using the properties of
only the most recent outburst}, rather than by estimating the
accretion history over the past \ee{4} yr, as is the case in the
core-dominated transients such as \aql\ and \cenx4\
\cite{rutledge01,rutledge01b}.  Our prediction of the quiescent
luminosity from the cooling crust following the outburst agrees, to
within observational and theoretical uncertainties (a factor of a
few), with the estimate of the observed bolometric luminosity.  Unlike
the case of the short $\tau_{\rm outburst}$ transients where we can
directly infer the average core temperature, in the case of \ks\ we
cannot measure the core temperature, and hence the recurrence time,
until after the crust has thermally relaxed (1--30~yr after the end of
the outburst).  Assuming that the source has been transiently
accreting for the past $10^{4}{\rm\,yr}$ or so, with outburst fluences
comparable to that of the most recent outburst, we can constrain the
core temperature to be less than the peak crust temperature, which,
using the relation of \acite{potekhin97} for an iron crust (low
conductivity), gives $T_{\rm core}\lesssim 2\tee{8}{\rm\,K}$.

We predict future lightcurves for \ks\ on the basis of 4 different
scenario for crustal and core microphysics (high/low conductivity,
and enhanced/normal neutrino emissivity).  These can be distinguished
from each other observationally -- one in $<$1 yr (excluding or
confirming the high $K$ crust with enhanced neutrino emissivity core),
others over longer terms (10-100 yrs).  This offers the unique
opportunity to directly constrain the properties of the NS crust --
thermal conductivity, heat capacity, and depth to the NS core --
through monitoring observations that would measure the predicted
decaying lightcurve.

\ks\ is now the fifth---after \cenx4, \aql, 4U 1608-522, and 4U
2129+47---field transient neutron star which displays in quiescence a
spectrum consistent with a H atmosphere NS of radius \rinfty$\sim$ 12
km, and (possibly) a hard ($\alpha\sim$0.85) power-law component which
comprises 10--40\% of the quiescent (0.5--10~keV) flux.  \ks\ shows no
variability on short ($<$hours) timescales.  These spectral and
variability properties are identical to those of the well-studied
\aql\ and \cenx4\ \cite{rutledge01,rutledge01b}, and the spectral
properties are similar to those from lower S/N data of 4U~1608-522 and
4U~2129+47 \cite{rutledge99,rutledge00}.  The derived radius is also
comparable to that derived from this object using the independent
method of a cooling radius expansion burst \cite{smith97}.

It is possible that a persistent source can turn off after long
periods of accretion (here, a persistent source is one accreting for a
period $\gg$\ee{4} yr, the core heating timescale).  While Eq.~1 would
seem to be applicable to predict the post turn-off quiescent thermal
luminosity from the crust, the equation assumes that all heat
deposited in the crust is re-radiated during quiescence.  That will
not be the case for sources approaching Eddington luminosity, such as
\ks, where the effect of neutrino cooling in the crust and core cannot
be neglected.  In such cases, $L_q \approx Q_{\rm crust} (\dot{M}_{\rm
outburst}/m_p) = 9\times 10^{33} (Q_{\rm crust}/150\,{\rm
eV})(\dot{M}_{\rm outburst}/10^{-9}\,M_\odot\,{\rm yr^{-1}})$.  Here
$Q_{\rm crust}$ is the energy per accreted baryon emitted through the
atmosphere from the crust immediately following the outburst.  A
minimum value, $Q_{\rm crust}=0.1 Q \approx 150$ eV/nucleon, applies
when the source accretes persistently at the Eddington limit, in which
only 10\% of the deposited heat flows towards the surface
\cite{brown00}.  At lower persistent luminosities, however, the crust
and core temperatures are lower and neutrino cooling in each becomes
less important, so $Q_{\rm crust}$ increases. In the case of \ks\
($L_{\rm outburst}\sim 0.1 L_{\rm Edd}$), we obtain $Q_{\rm
crust}\approx$250 eV, if \ks\ were a persistent accreter.  However,
the quiescent luminosity of \ks\ gives instead $Q_{\rm crust}=20$ eV,
a factor of 7 below the lower limit, even for a persistent source
accreting at the Eddington rate.  This demonstrates that \ks\ cannot
be a persistent accreter, because its crust (and by extension the
core) is too cool to have been accreting for $\gg$\ee{4} yr.  Note,
however, that we have neglected the effect of of neutrino emission due
to Cooper pairing of superfluid particles (see \citenp{yakovlev01} for
a review) which may reduce the value of $Q_{\rm crust}$.  

As discussed by \acite{rutledge00}, the study of qNSs in globular
clusters has yielded a new way to accurately measure NS radii.  The
distances to some globular clusters have been determined to \approxlt
2\% post-Hipparcos \cite{carretta00}, and can be measured even more
directly with SIM \cite{sim}, effectively removing distance as an
uncertainty.  At present, there is only one globular cluster qNS in
which the thermal component has been spectroscopically analyzed (in
$\omega$~Cen; \citenp{rutledge01c}) and there are two more proposed
(in 47 Tuc; \citenp{grindlay01}).  Measuring NS radii to an accuracy
of \ppm0.5~km---even without knowing the mass---can exclude $\sim$50\%
of the proposed equations of state for the NS core matter
\cite{lattimer01}.  Even though we presently cannot make such accurate
radius measurements in the field transients due to uncertain distances
(which can be overcome with SIM), they remain important targets for
detailed studies of the observational phenomena of qNSs, such as we
report here.

The relative luminosity of the (possible) power-law component poses a
puzzle.  It has been suggested that this component is due to accretion
onto the NS magnetosphere \cite{campana98b}.  However, such accretion
would be unrelated to the quiescent luminosity of NSs as predicted
from deep crustal heating (which is dominated by  outburst
accretion), and it can only be ascribed to coincidence that in the
cases where adequate sensitivity is obtained, the power-law component
produces a comparable fraction of the thermal flux in different
sources.  This suggests that the power-law component and H atmosphere
spectral component are perhaps more closely related than previously
thought.

\acknowledgements

The authors are grateful to the \chandra\ Observatory team for
producing this outstanding Observatory, and to the \chandra\ data
processing team who pre-handle all the data and provide the
calibrations which are used in this work.  We also thank
A. Y. Potekhin for reading and for comments on the text prior to
submission.  This research has made use of data obtained through the
High Energy Astrophysics Science Archive Research Center Online
Service, provided by the NASA/Goddard Space Flight Center; and of the
NASA/IPAC Infrared Science Archive, which is operated by the Jet
Propulsion Laboratory, California Institute of Technology, under
contract with the National Aeronautics and Space Administration. This
research was partially supported by the National Science Foundation
under Grant No. PHY99-07949 and by NASA through grants NAG 5-8658, NAG
5-7017, NAG 5-10865 and the \chandra\ Guest Observer program through
grant NAS GO0-1112B.  L. B. is a Cottrell Scholar of the Research
Corporation.  E. F. B. acknowledges support from an Enrico Fermi
Fellowship.  G. U. is a Lee A. DuBridge Fellow.

\clearpage

\begin{figure}[htb]
\caption{ \label{fig:ks1731} The $\nu F_\nu$ model spectrum of KS
1731-260, and the observed \chandra/ACIS-S BI data.  The solid line is
the best-fit unabsorbed spectrum (H atmosphere plus power-law, with
\rinfty =12.5 km and $\alpha$=0.85 held fixed) . The dashed line is the
H atmosphere component, the dashed-dotted line is the power-law
component.  The crosses are the observed \chandra\ data, with error
bars in countrate.  The two spectral components are equal near 3.5
keV, above which the power-law component dominates, and below which
the H atmosphere component dominates.
}
\end{figure}

\begin{figure}[htb]
\caption{ \label{fig:tevol} Crustal temperature distribution as 
a function of time, for $\tau_{\rm outburst}=13 yr$,
$3\times10^{-9}M_\odot$~yr$^{-1}$, recurrence time of 1500~yr,   and
deep crustal heating of $Q=1.45$~MeV, using ``low $K$''
conductivity, and modified Urca neutrino emission from the core.  
The $x$ axis is the density, $y$ axis is time since the beginning of
the outburst, and $z$ axis is temperature.
}
\end{figure}

\begin{figure}[htb]
\caption{ \label{fig:greg} Left panel {\bf (a)}: time evolution of the
quiescent luminosity assuming an outburst accretion rate of
$3\times10^{-9}M_\odot$~yr$^{-1}$, recurrence time of 1500~yr, and
deep crustal heating of $Q=1.45$~MeV/accreted baryon.  Solid and
dotted lines are for low crustal conductivity (set by
electron-impurity scattering), dashed and dash-dotted lines are for
high crustal conductivity (electron-phonon scattering).  Solid and
dashed line correspond to simulations with standard neutrino cooling
processes (modified Urca suppressed by nucleon superfluidity), dotted
and dash-dotted lines -- enhanced neutrino emission due to a core
$\pi$~condensate (with superfluid correction factors).  Right panel
{\bf (b)}: same as {\bf (a)}, except the nuclear energy release
$Q_{\rm nuc}$ in the crust is adjusted to give
$L_q\approx2.7\times10^{33}$~erg~s$^{-1}$ just following the outburst. 
$Q_{\rm nuc}$ values are: solid line --- $0.42$~MeV, dotted line ---
$0.49$~MeV, dashed line --- $0.6$~MeV, dash-dotted line ---
$3.1$~MeV.}
\end{figure}

\clearpage
\pagestyle{empty}
\begin{figure}[htb]
\PSbox{f1.eps hoffset=-80 voffset=-80}{14.7cm}{21.5cm}
\FigNum{\ref{fig:ks1731}}
\end{figure}

\clearpage
\pagestyle{empty}
\begin{figure}[htb]
\PSbox{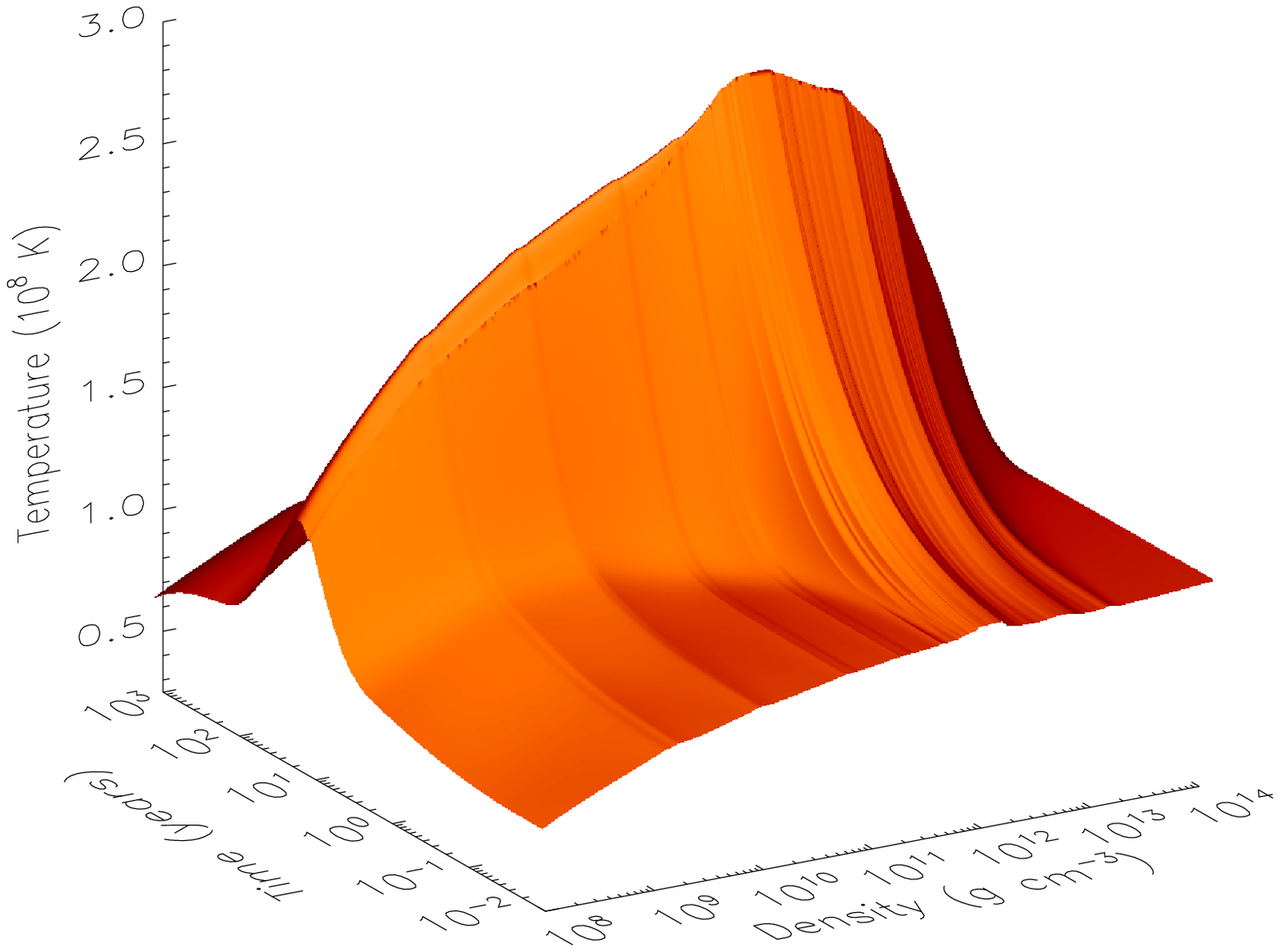 hoffset=-80 voffset=180}{14.7cm}{21.5cm}
\FigNum{\ref{fig:tevol}}
\end{figure}

\clearpage
\pagestyle{empty}
\begin{figure}[htb]
\PSbox{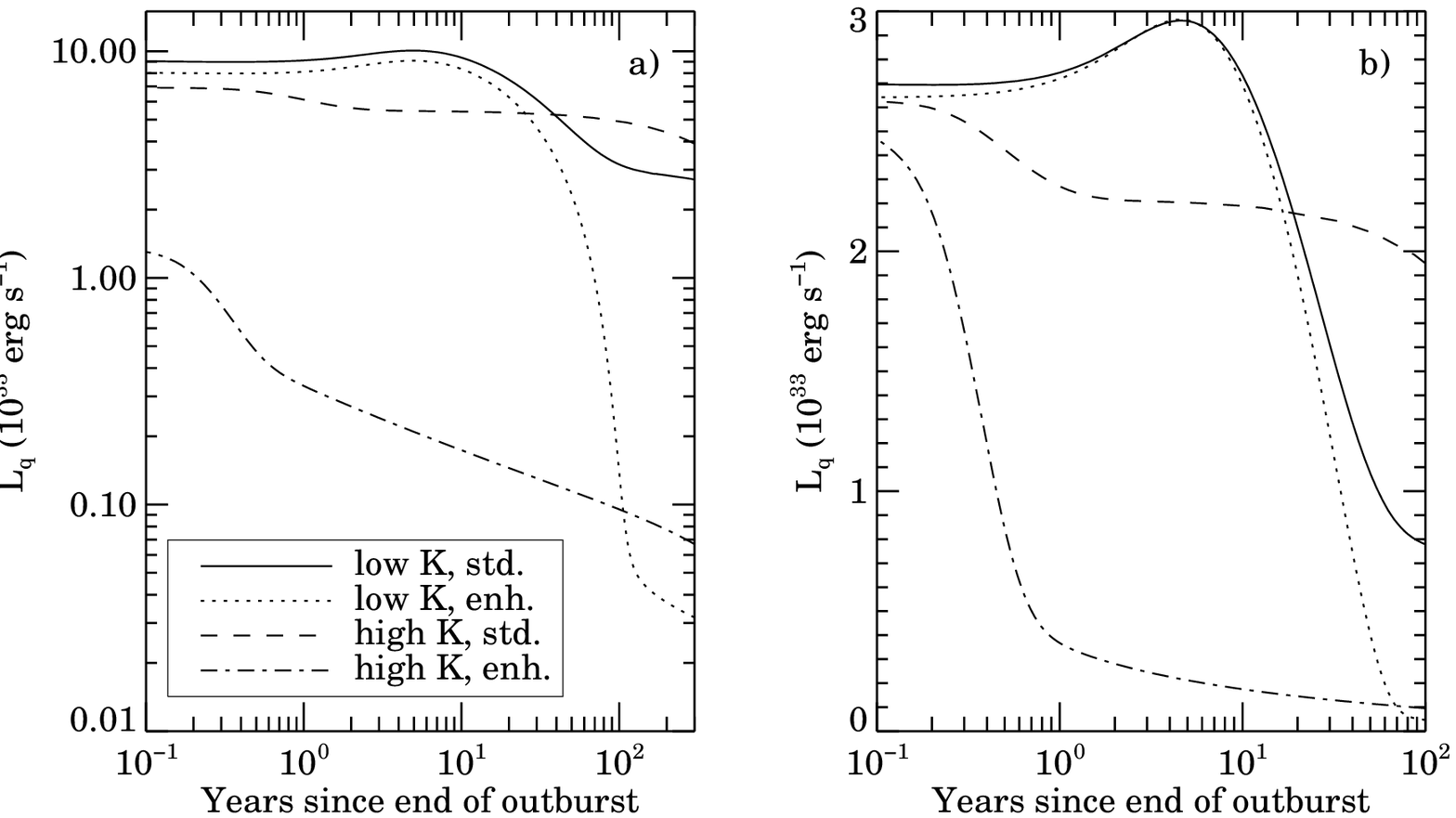 hoffset=0 voffset=180}{14.7cm}{21.5cm}
\FigNum{\ref{fig:greg}}
\end{figure}

\clearpage

\begin{deluxetable}{lllr}
\tablecaption{\label{tab:objs} Detected X-ray Sources in the field of
\ks }
\tablehead{
\colhead{} & 
\colhead{R.A.} &
\colhead{Dec.} &
\colhead{Counts$^a$}\\
\colhead{No.} & 
\colhead{(J2000)} &
\colhead{(J2000)} &
\colhead{} \\
}
\startdata
1 (\ks )		& 17h34m13.45s   & $-$26d05m18.7s & 180\ppm13\\
2 		& 17h34m12.68s	& $-$26d05m48.3s & 50\ppm7 		       \\ \hline		
\enddata
\tablecomments{Positional
uncertainty (from systematics of spacecraft pointing; http://cxc.harvard.edu/mta/ASPECT/celmon/) is
$\sim$1\arcsec. ~~~$^a$ Total background subtracted counts in 19401 sec
integration. }
\end{deluxetable}

\newcommand{\LH}{2}
\newcommand{\SingleSpace}{
  \renewcommand{\LH}{0.90}
  \def\baselinestretch{\LH}
  \tiny
  \normalsize
}

\begin{deluxetable}{lr}
\tiny
\tablecaption{\label{tab:fits} \chandra\ Spectral Model Parameters of
\ks\ (0.5-10 keV)}
\tablewidth{8cm}
\tablehead{
\colhead{Parameter} & 
\colhead{Value} \\
}
\startdata
\cutinhead{H Atmosphere}\\
\nhtt		& 1.0\ud{0.3}{0.2}	\\
\kteffinfty (eV) & 120\ppm30 \\
\rinfty\ (km)	 & 6.5\ud{6}{3}\\
Total Model Flux &  1.8		\\
\chisqrnu/dof (prob) & 1.06/7 (0.39)	\\
\cutinhead{H Atm.  + Power Law}\\
\nhtt		& 1.3\ppm0.3	\\
\kteffinfty (eV)& 90\ud{40}{20}\\
\rinfty\ (km)	& 23\ud{30}{15}	\\
$\alpha$	& -0.1\ud{1.4}{2.0}\\
$F_{X, PL}$	&  0.2		\\
Total Model Flux &  3.1  \\
\chisqrnu/dof (prob) & 0.43/5 (0.83)	\\
\cutinhead{H Atm. + Power Law (\rinfty, $\alpha$ fixed)}\\
\nhtt		& 1.06\ppm0.08	\\
\kteffinfty (eV)& 111\ppm3 		\\
\rinfty\ (km)	& (12.5) \\
$\alpha$	& (0.85)			\\
$F_{X, PL}$	&  0.2 	\\
Total Model Flux &  2.1 \\
\chisqrnu/dof (prob) & 0.47/7 (0.86)	\\
\cutinhead{Photon Power Law}\\
\nhtt		& 1.7\ppm0.4\\
$\alpha$  	& 5.2\ppm0.6\\
Total Model Flux &  20.0	\\
\chisqrnu/dof (prob) & 0.94/7 (0.48) \\
\cutinhead{Raymond-Smith}\\
\nhtt		& 0.7\ud{0.2}{0.1} \\
$Z$ $(Z_{\rm sol}$) & (1) \\
$kT$ (keV)	& 1.4\ppm0.1 \\
$\int n_e\, n_H dV$& (6.2\ppm1.5)\tee{55}\\
Total Model Flux &  0.93 \\
\chisqrnu/dof (prob) & 2.32/7 (0.02) \\
\cutinhead{Multicolor Disk}\\
\nhtt			& 1.1\ppm0.2\\
$T_{\rm in}$ (eV) 	& 370\ppm50 \\
$R_{\rm in}\sqrt{\cos(\theta)}$ (km)& 0.7\ud{0.5}{0.3}	\\
Total Model Flux	& 2.0		\\
\chisqrnu/dof (prob) & 1.1/7 (0.37)	\\
\cutinhead{Blackbody}\\
\nhtt			& 0.90\ppm0.2 \\
\kteffinfty\ (eV)  & 300\ppm 40 \\
\rinfty\ (km)	 &  1.3\ud{0.6}{0.4}\\
Total Model Flux & 1.3		\\	
\chisqrnu/dof (prob) & 1.11/7 (0.35) \\
\enddata 
\tablecomments{X-ray fluxes are un-absorbed, in units of \ee{-13}
\cgsflux\ (0.5-10 keV).  Uncertainties are 1$\sigma$. Values in
parenthesis are held fixed.  Assumed source distance d=8 kpc.}
\end{deluxetable}

\end{document}